\documentstyle[sprocl]{article}
\def\Journal#1#2#3#4{{#1} {\bf #2}, #3 (#4)}
\def\ANP{\em Ann. of Phys. (NY)}
\def\NCA{\em Nuovo Cimento}
\def\NPA{{\em Nucl. Phys.} A}
\def\NPB{{\em Nucl. Phys.} B}
\def\PLB{{\em Phys. Lett.}  B}
\def\PRL{\em Phys. Rev. Lett.}
\def\PRR{{\em Phys. Rev.}}
\def\PRB{{\em Phys. Rev.} B}
\def\PRD{{\em Phys. Rev.} D}

\def\be{\begin{equation}}
\def\ee{\end{equation}}
\def\bea{\begin{eqnarray}}
\def\eea{\end{eqnarray}}
\begin{document}
\title{COOL PIONS MOVE AT LESS THAN THE SPEED OF LIGHT} \author{ROBERT
D. PISARSKI AND MICHEL TYTGAT} 
\address{Dept. of Physics, Brookhaven
National Laboratory,\\ NY 11973, USA} 
\maketitle
\abstracts{ At nonzero
temperature, pions propagate through a thermal medium at less than the
speed of light.  About low temperature, this effect begins not at
$\sim T^2$, but at next to leading order, $\sim T^4$.  We also derive
the generalization of the relation of Gell-Mann, Oakes, and Renner to
nonzero temperature.}

\section{General analysis}

In this proceeding we give a pedagogical explanation of recent work of
ours.~\cite{pistyt} In this section we show how the velocity of pions,
or more generally Goldstone bosons, changes with temperature in a
thermal distribution.

This in itself is pretty trivial, and familiar from other contexts.
For examples, the appropriate analogy to Goldstone bosons in a theory
which is relativistically invariant at zero temperature are spin waves
in an antiferromagnet.~\cite{anti,leutnon} This is because the
dispersion relation for such spin waves is $\omega = v p$, where
$\omega$ is the frequency, $p$ the magnitude of the spatial momentum,
and so $v$ is the velocity.  (Spin waves in a ferromagnet behave as
$\omega \sim p^2$.)  In a relativistically invariant system, of course
$v=c$, but in antiferromagnets, the velocity $v$ is just some
parameter that depends upon details of the couplings, etc.  Thus for
antiferromagnets, it is completely unremarkable that $v$ changes with
temperature, since the couplings change with temperature as well.

This is in agreement with Goldstone's theorem, which tells us that for
the Goldstone modes, the inverse pion propagator vanishes at zero
spatial momentum, $p=0$.~\cite{goldstone} Thus it is fine if $\omega$
vanishes like some constant $v$ times $p$, where $v$ changes with
temperature.

Less trivially, one can also understand the damping of pions or spin
waves.~\cite{anti} Even without calculation, we expect that the
damping rate vanishes at zero spatial momentum, since by Goldstone's
theorem the inverse pion propagator must vanish at zero momentum, for
both the real and imaginary parts.  How precisely the imaginary parts,
and so the damping rate, vanishes is an interesting question which we
face when we get into the gory details below.

A similar problem is the propagation of light in a medium with an
index of refraction greater than one.~\cite{latorre} The analogy is
imprecise, though, because in a thermal bath Debye screening generates
a finite correlation length for photons, whereas in the broken
symmetric phase, Goldstone bosons always have an infinite correlation
length.

A better example is the propagation of light in a background magnetic
field, where at two loop order the velocity of light is less than that
in vacuum by an amount $\sim \alpha^2 B^2/m^4$, where $m$ is the mass
of the electron.~\cite{adler} There is a technical similarity here as
well.  The deviations from the speed of light are proportional to the
energy density $\sim B^2$ for photons in a background $B$ field, and
to the free energy density $\sim T^4$ for thermal pions.  For the
former, this is unsurprising, since one wouldn't expect anything to
depend upon $B$, only on $B^2$.  But for thermal pions, that there
aren't terms of $\sim T^2$ is at first sight unexpected.  However old
arguments in a different guise~\cite{dey} tell us that to leading
order, $\sim T^2$, lorentz covariance is manifest in a pion gas, so
$v=1$.  Indeed, it is amusing that for thermal pions, the deviation of
$v^2$ from unity is proportional to $8 \pi^2/45\, T^4 $,
Eqs. (\ref{eq:fa}) and (\ref{eq:fd}), which is $16/3$ times the free
energy density of a relativistic pion gas.

Quite remarkably, this is analogous to the behavior of spin waves in
ferromagnets at low temperature, where the dispersion relation is
$\omega(k) = c(T) k^2$. As first shown by Dyson \cite{dyson}, $c(T)$
changes with temperature but only at next to leading order in an
expansion about $T=0$ and the correction is proportional to the free
energy density of the gas of spin waves. Presumably, the same is true
for other systems with Goldstone modes at low temperature. Also,
perhaps this a hint of the existence of a more general relation, which
is valid nonperturbatively; analogous, say, to the expression for the
speed of sound in a medium in terms of the pressure and density.

Consequently, the basic physics which we consider is hardly
revolutionary.  Nevertheless, we think that we have a
novel way of looking at these phenomena, for instead of considering
inverse (pion) propagators, we use $PCAC$, and speak of pion
decay constants.  

At zero temperature, the pion decay constant
$f_\pi \sim 93 \, MeV$ is defined by 
\begin{equation}
\langle 0 \vert A^\mu_{a} \vert \pi^b(P) \rangle 
= i f_\pi \delta^{ab} P^\mu \; ,
\label{eq:ea}
\end{equation}
where $A^\mu_{a}$ is the axial vector current, 
and the pion has momentum
$P^\mu = (p^0,\vec{p})$.

Once stated, our basic point is obvious.
At nonzero temperature, because of the presence of the
medium, there are, in general, {\it two} distinct
pion decay constants.  The timelike component of the current has one,
$$
\langle 0 \vert A^0_{a} \vert \pi^b(P) \rangle_T 
= i f^t_\pi \delta^{ab} p^0 \; ,
$$
and the spatial part of the current has another,
\begin{equation}
\langle 0 \vert A^i_{a} \vert \pi^b(P) \rangle_T 
= i f^s_\pi \delta^{ab} p^i \; .
\label{eq:eb}
\end{equation}

At this point we should be careful and qualify exactly in which
regime we are computing.  First, all matrix elements
are computed at a temperature $T$ in the imaginary time formalism;
then, the timelike component
of the momentum, $p^0$, is analytically continued from
euclidean to Minkowski values, $p^0 = - i \omega + 0^+$.  
The two pion decay constants, 
$f_\pi^t$ and $f_\pi^s$ are defined
about zero momentum, $\omega$ and $p \rightarrow 0$; typically,
each has both a real and an imaginary part.

In a colloquial fashion, we only deal with soft, ``cool'' pions.
Soft means that all components of the momenta are small relative
to (the real parts) of $f^t_\pi$ and $f^s_\pi$.
``Cool'' means that the temperature is well below that for the
restoration of chiral symmetry.  This is certainly a well defined
regime, where the sigma meson is heavy and its 
abundance is Boltzmann suppressed.
Once the sigma becomes light, with a mass less than the $f_\pi$'s, then
one enters the regime of the critical point, and our analysis
does not apply.

We certainly aren't the first people to realize that
there are two pion decay constants at nonzero temperature:
that's an old story in nonrelativistic systems;~\cite{leutnon}
for pions, it was pointed out before by 
Kirchbach and Riska,~\cite{kirchbach} by
Thorsson and Wirzba,~\cite{thorsson} and probably by others as well.

What we want to emphasize is  that whenever there are two, distinct
pion decay constants, then the propagation of pions {\it must}
move off of the light cone.  The argument is elementary: whether
or not the vacuum respects the chiral symmetry, the symmetry
is still good {\it on} the pion mass shell.  Thus if we take
the derivative of the axial current, we can use it to {\it define}
the pion mass shell:
\begin{equation}
f_\pi^t \, p_0^2 + f_\pi^s \, p^2 = 0|_{\pi \; mass \; shell}  \; .
\label{eq:ec}
\end{equation}
This expression has both a real and an imaginary part.  
The mass shell condition for pions is then,
\begin{equation}
p^0 = - i \omega - \gamma \; ,
\end{equation}
where the real part gives
\begin{equation}
\omega^2
= v^2 p^2 \approx
\frac{\mbox{Re} \, f_\pi^s}{\mbox{Re} \, f_\pi^t} \; p^2 \; .
\label{eq:ed}
\end{equation}
and the imaginary part fixes 
\begin{equation}
\gamma \approx \frac{1}{2 \omega \, \mbox{\mbox{Re}}\, f_\pi^t}
\left( + \mbox{Im} \, f^t_\pi \; \omega^2 - \mbox{Im} 
\, f^s_\pi \; p^2 \right) \geq 0 \; ,
\label{eq:eab}
\end{equation}
In doing so, we assume
that the imaginary parts are small relative to
the real parts,
\begin{equation}
\mbox{Im} \, f_\pi^{t,s} \ll \mbox{\mbox{Re}}\, f_\pi^{t,s} \; .
\label{eq:eea}
\end{equation}
One could probably gain some general understanding of what happens
when this condition doesn't hold, and pions are strongly damped,
but we haven't tried.

With Eq. (\ref{eq:eab}) in hand, we can understand the possible
behavior of the damping rate about zero momentum.  The
simplest possibility is that the imaginary parts of
$f_\pi^t$ and $f_\pi^s$ are nonzero at zero momentum; then
the damping rate vanishes linearly as $p \rightarrow 0$.
One can show that in this case, the imaginary part of the
pion self energy vanishes quadratically about zero momentum,
so Goldstone's theorem is evidently obeyed. However,
in some  cases in which calculations have been done,
the damping rate vanishes faster than linearly.  For example,
in the linear sigma model discussed in the next section,
it vanishes exponentially.  This is special to the linear sigma
model at the order treated, since the only intermediate state
through which the pion can scatter involves a virtual sigma
meson, which is heavy, and so Boltzmann suppressed.  Also,
for spin waves in an antiferromagnet, the damping rate vanishes
quadratically about zero momentum.~\cite{anti}  

We quote, without derivation, our results~\cite{pistyt} on
the generalization of the relation of Gell-Mann, Oakes, and Renner
to nonzero temperature.  Using methods of Shore and Veneziano,~\cite{shore} 
we find that the dynamic pion mass, which is
the pole for complex $p^0$ at $p=0$, is
\begin{equation}
m^2_\pi = \frac{2 m \langle \overline{q} q\rangle_T}
{(\mbox{Re} \, f_\pi^t)^2} \; .
\label{eq:ek}
\end{equation}
This is the same expression as at zero temperature, except that
instead of $f_\pi$ entering, it is the real part of $f_\pi^t$ which
appears.  Of course, both $\mbox{Re} \, f_\pi^t$ and the value of the
quark chiral condensate, $\langle \overline{q} q\rangle_T$, are
functions of temperature.  A relation like this was first obtained by
Thorsson and Wirzba,~\cite{thorsson} although they just wrote
$f_\pi^t$ instead of its real part.

It is worth emphasizing that this is the dynamic pion mass,
and not the static pion mass.  The latter is given by the
pole for zero frequency, $p^0=0$, in the complex $p$ plane.
This mass is just 
\begin{equation}
m_\pi^{static} = \frac{m_\pi}{v} \; ,
\end{equation}
and so is less than $m_\pi$ when $v < 1$.  This is interesting,
because what is typically measured in numerical simulations
of lattice gauge theory are the static and not the dynamic masses;
the static mass is then an upper bound on the dynamic mass.

\section{Lowest order}

We now consider where these effects first appear in
an expansion about zero temperature.  Using either
a nonlinear~\cite{gassert} or a linear~\cite{bochkarev}
sigma model, to leading order in $T^2/f_\pi^2$,
$f_\pi^t(T) = f_\pi^s(T)$, so pions move 
undamped at the speed of light.
The reason for this was given by Dey, Eletsky, and Ioffe,~\cite{dey}
who considered the thermal average of the two point function of
either vector or axial vector currents.
For completeness, we redo their analysis in terms of the
thermal average of the one point function\footnote{Strictly speaking, 
the Gibbs average is only defined for diagonal
matrix elements but, to leading order in $T^2$ at least, (\ref{thtr}) 
reproduces the
known answer. It would be interesting to know what is the
definition valid to any order.},
\begin{equation}
\label{thtr}
\langle 0 \vert  A^\mu_a \vert \pi^b\rangle_T \, 
\hat = \, {\sum_{n} \langle n \vert e^{-H/T} A^\mu_a 
\vert n; \pi^b \rangle \over 
\sum_{n} \langle n \vert e^{-H/T}\vert n\rangle}
\end{equation}
At low temperature, $T \ll f_\pi$, the states 
$\vert n\rangle$ contain only pions.  Further, 
to lowest order in $T^2/f_\pi^2$ we can truncate
states with the fewest number of pions to obtain
\begin{equation}
\langle 0 \vert A^\mu_a \vert \pi^b\rangle_T = \langle 0 \vert A^\mu_a
\vert \pi^b\rangle_{T=0} + \; \sum_c \int {d^3k \over (2\pi)^3}
{1\over 2 k} n(k) \langle \pi^c(k) \vert A^\mu_a \vert \pi^c(k);
\pi^b\rangle_{T=0} + \ldots
\end{equation}
where $k = \vert {\bf k}\vert$ and $n(k)$ is the Bose-Einstein 
statistical distribution function.  The Bose-Einstein distribution
function is built up by summing over states: in the bra state,
first one pion with momentum $k$, weighted as $e^{-k/T}$,
then two pions with the same momentum, weighted as $e^{-2k/T}$,
{\it etc.}  In bra states with more than one pion, all pions
beyond the first
are completely disconnected from the matrix element, and only
contribute through their statistical weighting.  Summing
over all such pions then generates $e^{-k/T}/(1-e^{-k/T})
= n(k)$.
Using $A^a_0$ as an interpolating field for $\pi^a$,
and the canonical commutation relations of current algebra,
the thermal pions $\pi^c(k)$ are eliminated to obtain
\begin{equation}
\sum_c\langle \pi^c(k) \vert A^\mu_a \vert \pi^c(k); 
\pi^b\rangle_{T=0} \sim - {2 \over f_\pi^2}\langle 0\vert 
A^\mu_a \vert \pi^b\rangle_{T=0} \; .
\end{equation}
To leading order in $T^2/f_\pi^2$, then, 
\begin{equation}
\langle 0 \vert  A^\mu_a \vert \pi^b\rangle_T 
= \left(1 - {T^2\over12 f_\pi^2}\right)\langle 0 \vert  
A^\mu_a \vert \pi^b\rangle_{T=0} . 
\end{equation}
Since the matrix element is lorentz covariant at zero temperature,
it remains so to $\sim T^2/f_\pi^2$, and $f_\pi^t = f_\pi^s$.

\section{Nonlinear sigma model}

It would of great interest to perform calculations in the nonlinear
sigma model, to ask: how does the damping rate vanish about
zero momentum? and, is $v<1$?  Both effects can first
appear at two loop order, $\sim T^4/f_\pi^4$.
Calculations of the damping rate have been performed by
Gavin {\it et al.};~\cite{non2,schenka,schenkb} however,
they used various (physically reasonable) approximations
to the $\pi\pi$ scattering amplitude, and so it is not apparent
how $\gamma(p)$ vanishes as $p \rightarrow 0$ in the chiral
limit \footnote{To leading order in a virial expansion, as eq. (2.4)
of ref. [11], we estimate  $\gamma \sim p \, (T^4/f_\pi^4)$
about zero momentum in the chiral limit.}. 

We note, however, that there is indirect support for $v<1$ in
the results of Schenk.~\cite{schenka,schenkb}  
He also computed not in the strict chiral limit, but using various
approximations to the $\pi\pi$ scattering amplitude.
In fig. 5 of Ref. [11] and fig. 7 of Ref. [12], Schenk graphs
$f(p) = \omega(p)/\sqrt{p^2 + (m_\pi^0)^2}$, where $m_\pi^0$ is the
mass at zero temperature, and $\omega(p)$ his computed quasiparticle
pion mass.

At leading order,
$\omega_{lo}(0) > m_\pi^0$, and so $f_{lo}(0) > 1$.  As $p$ increases,
$f_{lo}(p)$ decreases monotonically to one.  
This is what happens when
$\omega_{lo}(p) = \sqrt{p^2 + \omega_{lo}(0)^2}$, or $v_{lo}=1$.

At next to leading order, $\omega_{nlo}(0) < m_\pi^0$.  This is not
particularly suprising.  However, as $p$ increases, $f_{nlo}(p)$ first
{\it decreases}, and then increases, approaching one from below.
That is, there is a ``dip'' in $f_{nlo}(p)$.  Assume that the
dispersion relation is
\begin{equation}
\omega_{nlo}(p)^2 = v_{nlo}(p)^2 p^2 + \omega_{nlo}(0)^2 \; .
\end{equation}
Of course $v_{nlo}(p) \rightarrow 1$ 
for $p \gg \omega_{nlo}(0)$.  For the dip to
occur, however, at zero momentum the velocity must satisfy
\begin{equation}
v_{nlo}(0) < \frac{\omega_{nlo}(0)}{m_\pi^0} < 1 \; .
\end{equation}
The first inequality guarantees that a dip occurs 
in $f_{nlo}(p)$; the second
is because Schenk's results show that $\omega^{nlo}(0) < m_\pi^0$.
These are both strict inequalities, and so $v_{nlo}(0) < 1$.
This is in accord with our analysis.

\section{Linear sigma model}

In this section we describe, 
hopefully in a more comprehensible way than in Ref. [1], where these effects
first show up in a linear sigma model in an expansion about
zero temperature.  The modifications of pion propagation were
already computed long ago by Itoyama and Mueller;~\cite{itoyama}
the only thing which we are doing is to turn their results
for the pion propagator into results for $f_\pi^t$ and $f_\pi^s$.

The lagrangian of the linear sigma model is given by
\begin{equation}
{\cal L} = \frac{1}{2} \left( \partial_\mu \phi \right)^2
- \frac{\mu^2}{2} \phi^2 
+ \frac{\lambda}{4} \left(\phi^2 \right)^2 - h \sigma \; ,
\label{eq:eo}
\end{equation}
where $\phi = (\sigma,\vec{\pi})$ is an $O(4)$ isovector field.
We introduce a background magnetic field $h$ which is proportional
to the current quark mass $m$.  For $h=0$,
the vacuum expectation value of the $\sigma$ is 
$\sigma_0 = \sqrt{\mu^2/\lambda}$, where we then shift
$\sigma \rightarrow \sigma_0 + \sigma$; 
for two flavors, $f_\pi = \sigma_0$.

We want to compute terms of order $\sim T^4$.  There are many such
terms, both at one and at two loop order.  To avoid being drowned
in calculational details, we extract a small subset of terms
which are dominant in weak coupling, those 
$\sim T^4/(f_\pi^2 m_\sigma^2)$ at one loop order.  
In weak coupling, $m_\sigma^2 = 2 \lambda f_\pi^2$,
and so these terms are larger by $1/\lambda$ than terms
$\sim T^4/f_\pi^4$.  Indeed, we are lucky not to have to
go to two loop order in the first place, as is required in the
nonlinear model.

We first review the calculations of Itoyama and Mueller.~\cite{itoyama}
There are several diagrams which contribute to the pion self energy,
fig. 5 of Ref. [13].  However, the only one which can produce
the effects we are looking for is where a pion splits into a pion
and a sigma, and then recombines.  This is because we need a 
nontrivial momentum dependence in the pion self energy, and this
is the only diagram that has it.  The integral for this
particular diagram is proportional to
\begin{equation}
{\cal I}(P) = tr_K \frac{1}{K^2 ((P-K)^2 + m^2_\sigma) } \; ,
\label{eq:ep}
\end{equation}
where $tr_K = T \sum_{n = -\infty}^{+\infty} \int d^3 k/(2 \pi)^3$.

For the real parts of the pion self energy, the trick is simply to
expand this integral in powers of the external momentum $P$.
At lowest order, this is trivial, since we just take the
propagator for the sigma meson, and replace it by a constant:
\begin{equation}
{\cal I}(P) \sim 
\frac{1}{m_\sigma^2} \; tr_K \frac{1}{K^2} \sim 
\frac{1}{m_\sigma^2} \; \frac{T^2}{12} \; .
\label{eq:eq}
\end{equation}
We can forget about any ultraviolet divergences along the way,
since they will be taken care of by zero temperature renormalization,
as usual.  (We might have to worry if we found any logarithmic
ultraviolet divergences, but we're only throwing away powers divergences,
which is ok.)

To go beyond leading order, then, to $\sim T^4$, all we have to
do is to expand the integral in $P$ to $\sim P^2$.  Doing so,
we find that there are two types of integrals which enter.  One
is that of Eq. (\ref{eq:eq}), while the other is
\begin{equation}
tr_K \frac{K^\mu K^\nu}{K^2} \sim 
\left( \delta^{\mu \nu} - 4 n^\mu n^\nu \right)
\frac{\pi^2 T^4}{90} \; ,
\label{eq:er}
\end{equation}
where $n^\mu = (1,\vec{0})$.  This appearance of the timelike
vector $n^\mu$ is entirely where $v<c$ will enter at this order,
since it produces different terms in the pion self energy
$\sim p_0^2$ and $p^2$.  For both integrals in Eqs. (\ref{eq:eq})
and (\ref{eq:er}), evidently the scale of the loop momentum $K \sim T$.
This justifies the approximations made.

These are all the integrals which are needed to get the
real part of the pion self energy.  For the imaginary part,
we need the integral
\begin{equation}
\mbox{Im} \; {\cal I}(P)|_{\omega \sim p \ll m_\sigma}
\sim \frac{1}{16 \pi} \; 
exp \left( - \frac{m^2_\sigma}{4 p T} \right) \; .
\label{eq:eu}
\end{equation}
The imaginary part is exponentially small because one
needs to scatter off of a sigma meson in the thermal distribution
at low temperature.  Sigmas are heavy at low temperature, so 
the probability to pop one out of the distribution
is Boltzmann suppressed.  

Introducing the quantities 
\begin{equation}
t_1 = \frac{T^2}{12 f_\pi^2} \;\;\; , \;\;\;
t_2 = \frac{\pi^2}{45} \frac{T^4}{f_\pi^2 m_\sigma^2} \;\;\; , \;\;\;
t_3 = \frac{1}{32 \pi} \frac{m^4_\sigma}{f_\pi^2 p^2}
\; exp \left( - \frac{m^2_\sigma}{4 p T} \right) \; ,
\label{eq:fa}
\end{equation}
to this order the inverse pion propagator is
\begin{equation}
Z_\pi \Delta^{-1}(P) \sim
(1 + t_1 + 6 t_2) p_0^2
+ (1 + t_1 -  2 t_2) p^2
+ m^2_\pi \left( 1 + 3 t_1/2 \right) - 2 i p^2 t_3 \; .
\end{equation}
The zero of this expression determines the pion mass
shell.  In the chiral limit, it is
\begin{equation}
i p^0 \sim v p - i p \, t_3 \; ,
\label{eq:fd}
\end{equation}
where the pion velocity is given by 
\begin{equation}
v^2 \sim 1 - 8 t_2 \; ,
\label{eq:fc}
\end{equation}
The result for the pion velocity agrees with (A.8) of Ref. [13].
There appears to be a misprint for the pion damping rate
in (A.8') of Ref. [13].

To evaluate $f_\pi^t$ and $f_\pi^s$ in the linear sigma model 
we need the axial current,
\begin{equation}
A_\mu^{a} = (\sigma_0 + \sigma) \partial_\mu \pi^a 
- \pi^a \partial_\mu \sigma \; .
\label{eq:ex}
\end{equation}
Then we take the two point function of axial currents and
saturate it with a single pion in the intermediate state;
from Eqs. (9) and (11) of Ref. [1], this equals
\begin{equation}
\langle 0 | \partial^\mu A^a_\mu \partial^\nu A^b_\nu | 0\rangle_T
\sim \delta^{a b} ({\mbox Re} f_\pi^t ) \left( f_\pi^t p_0^2
+ f_\pi^s p^2 \right) \; ,
\label{eq:exx}
\end{equation}
Generally, for small momenta this two point function is dominated
by the contribution of a single pion, since the contribution of
other states is of higher order in momenta, $\sim (p^2)^2$, etc.
This remains true even if the condition of
Eq. (\ref{eq:eea}) is not satisfied:
even for strongly damped pions, the dominant contribution to
Eq. (\ref{eq:exx}) arises from single pions in the intermediate state.

At one loop order, the diagrams which contribute to $f^t_\pi$ and
$f^s_\pi$ are given in fig. (5) of Ref. [15].  As for the pion self
energy, the only diagrams which can produce the effects which we are
looking for are those involving a virtual pion-sigma meson loop.  In
the calculation of the two point function in Eq. (\ref{eq:exx}), this
kind of loop enters in two ways.  First, simply as self energy
insertions on the intermediate pion.  Secondly, from the form of the
axial current, they can also enter through either vertex, as a type of
form factor.

For the real parts of the pion decay constants, the integrals
in Eqs. (\ref{eq:eq}) and (\ref{eq:er}) suffice.  For the form
factor diagram involving a loop with a sigma meson and a pion,
again one expands the propagator of the sigma meson in powers
of the external momentum.  As a form factor, only terms linear
in $P$ need be retained.

For the imaginary parts, we need two new integrals, 
\begin{equation}
tr_K \frac{k^0}{K^2 ((P-K)^2 + m_\sigma^2)} \sim \frac{i}{16 \pi}
\left( \frac{m^2_\sigma}{4 p} + T \right) \; exp \left( -
\frac{m^2_\sigma}{4 p T} \right) \; .
\label{eq:ey}
\end{equation}
and
\begin{equation}
tr_K \frac{k^i}{K^2 ((P-K)^2 + m_\sigma^2)}
\sim \frac{p^i}{16 p \pi} \left( \frac{m^2_\sigma}{4 p} - T \right)
\; exp \left( - \frac{m^2_\sigma}{4 p T} \right) \; .
\label{eq:ez}
\end{equation}
These expressions are valid near the light cone,
$\omega \sim p \ll m_\sigma$.
In these integrals, the dominant loop momentum $k$ is very large,
$k \sim m^2_\sigma/p \gg m_\sigma$.  Thus one typically retains
only those terms $\sim k$ and not $\sim p$.  

The results for $f_\pi^t$ and $f_\pi^s$ are
\begin{equation}
f_\pi^t \sim  \left( 1 - t_1 + 3 t_2 + i t_3\right) f_\pi \;\;\; , \;\;\;
f_\pi^s
\sim \left( 1 - t_1 - 5 t_2 - i t_3\right) f_\pi \; .
\label{eq:fb}
\end{equation}
It is then simple to verify that Eqs. (\ref{eq:ed}) and (\ref{eq:eab})
give the same mass shell condition as found directly from the pion
propagator in Eq. (\ref{eq:fd}).  

Given that 
\begin{equation}
\langle \overline{q} q \rangle \sim \sigma_0(T) 
\sim \sigma_0(0) \left( 1 - 3 t_1/2 \right) \; ,
\label{eq:ff}
\end{equation}
we also verify our generalization of the formula of
Gell-Mann, Oakes, and Renner in Eq. (\ref{eq:ek}) for
the dynamic pion mass:
\begin{equation}
m^2_\pi(T)  \sim m^2_\pi \left(
1 + t_1/2 - 6 t_2 \right) \; .
\label{eq:fg}
\end{equation}
From the propagator, this result is really trivial, since 
it follows simply by using the wave function renormalization of the
pion, $Z_\pi \sim 1 + t_1 + 6 t_2$,
and finding the pole in $p_0^2$.  In detail, what matters is
that the coefficient of $t_2$ in $Z_\pi$ and in $(f_\pi^t)^2$ are
equal.

\section*{Acknowledgments}

R. D. P. would like to thank Misha Shifman for inciteful
comments made during a talk on a related subject in
November, 1995.  Misha began by stating that
``Using PCAC at nonzero temperature...'', and then went on from there.
This led us to pause and ask, exactly what {\it does}
PCAC mean at nonzero temperature?   This is the really the essence
of our work.

M.T. would like to thank R. Brout for his comments on spin waves in 
ferromagnets.

This work is supported by a DOE grant at 
Brookhaven National Laboratory, DE-AC02-76CH00016.

\end{document}